\documentclass[prl,aps,twocolumn,showpacs,superscriptaddress,floatfix]{revtex4-1}

\usepackage{amssymb}
\usepackage{amsmath}
\usepackage{graphicx,bbm,color}
\allowdisplaybreaks
\usepackage{verbatim}
\usepackage{relsize}
\usepackage{hyperref}

\usepackage{dsfont}

\newcommand{\tr}{\operatorname{tr}}

\usepackage{dsfont}
\newcommand{\bgt}{\begin{itemize}}
\newcommand{\ent}{\end{itemize}}

\newcommand{\op}{\operatorname}

\newcommand{\lan}{\langle}
\newcommand{\ran}{\rangle}

\newcommand{\Tr}{\operatorname{Tr}}

\newcommand{\E}{\op{\mathbb{E}}}

\newcommand{\bbm}{\begin{bmatrix}}
\newcommand{\ebm}{\end{bmatrix}}
\newcommand{\bes}{\begin{equation*}}
\newcommand{\ees}{\end{equation*}}
\newcommand{\be}{\begin{equation}}
\newcommand{\ee}{\end{equation}}
\newcommand{\beqy}{\begin{eqnarray}}
\newcommand{\eeqy}{\end{eqnarray}}
\newcommand{\beq}{\begin{eqnarray*}}
\newcommand{\eeq}{\end{eqnarray*}}

\newcommand{\bpm}{\begin{pmatrix}}
\newcommand{\epm}{\end{pmatrix}}

\long\def\symbolfootnote[#1]#2{\begingroup
\def\thefootnote{\fnsymbol{footnote}}\footnote[#1]{#2}\endgroup}

\renewcommand\Re{\operatorname{Re}}
\renewcommand\Im{\operatorname{Im}}

\begin{document}

\setkeys{Gin}{width=0.9\textwidth}
% scaling the size of pictures 

\title{Dynamical typicality of embedded quantum systems}

\author{Gr\'egoire Ithier}      \affiliation{Department of Physics, Royal Holloway, University of London, United Kingdom}
\author{Florent Benaych-Georges}     \affiliation{MAP5, UMR CNRS 8145 -- Universit\'e Paris Descartes, France}

%\pacs{85.25C.p}

\begin{abstract}
 We consider the dynamics of an arbitrary quantum system coupled to a large arbitrary and fully quantum mechanical environment through a random interaction. We 
 establish analytically and check numerically the typicality of this dynamics, in other words the fact that the reduced density matrix of the system has a self-averaging property.
% the self-averaging property of the reduced density matrix of the system, in other words, the fact that the dynamics of the system is universal
 % the \textit{typicality} of this dynamics, in other words, its universal character which is due to a self-averaging of the reduced density matrix of the system. 
 This phenomenon, which lies in a generalized central limit theorem, justifies rigorously averaging procedures over certain classes of
   random interactions and can explain the absence of sensitivity to microscopic details of irreversible processes such as thermalisation. It provides more generally a new ergodic principle for
    embedded quantum systems.
\end{abstract}

\maketitle

Thermalisation is probably one of the most common phenomenon in nature. Its   
\textit{universality}, i.e. the fact that it does not depend on microscopic details but only on a small set of macroscopic parameters  (like temperature or pressure) 
has been known experimentally for a long time~\cite{Carnot,Boltzmann1974}.
In typical conditions, non equilibrium dynamics is expected to lead to some stationary state, independent of initial conditions,
where macroscopic quantities can be calculated using statistical thermodynamics~\cite{Kubo,LaudauLifschitz,Gemmer}.
 Despite being broadly accepted, the foundations of this statistical framework are relying on a set of assumptions 
 where the role of randomness and the associated lack of knowledge, the role of averaging over this randomness and
the supposed link with temporal averages through ergodicity, are not justified in a satisfactory manner (see e.g. the discussion in~\cite{gemmer_distribution_2003}). 

On the various attempts for setting statistical mechanics on the firm ground of quantum theory, \textit{typicality} statements are one of the most promising.
They introduce some randomness in the problem and, relying on a key mathematical phenomenon:  ``measure concentration''~\cite{NoteMeasConc,MR1387624},
 % previously used in 
they show that surprisingly this randomness actually \textit{does not matter}, as soon as the Hilbert space dimension of the system considered is large enough.
Such randomness has been previously introduced on the choice of a global quantum state\cite{popescu_entanglement_2006,goldstein_canonical_2006,reimann_typicality_2007,linden_quantum_2009}
%riera_thermalization_2012,tasaki_quantum_1998,
 (i.e. a state of a large closed system)
and on the choice of the full Hamiltonian\cite{vonNeumann,reimann_typical_2016}, in order to get respectively a local property, like the state of a subsystem,
or some global property, like a macroscopic observable.
The former approach, which provides ``canonical typicality'' is purely kinematic in the sense that it is fundamentally a consequence of the geometry of the space of states, and does not provide any
 dynamical information. On the other hand the later approach (called nowadays ``normal typicality'', see \cite{goldstein_normal_2010,goldstein_long-time_2010}) is dynamical but by randomizing the full Hamiltonian, it does not allow to consider a specific system or environment.

In this Article, we consider the generic problem of a quantum system coupled to a quantum environment where we introduce randomness at the level of the interaction Hamiltonian \textit{only}. By doing so we get dynamical results 
for several classes of interaction Hamiltonians and most importantly for \textit{arbitrary} system, environment, and \textit{global} initial state (i.e. of system and environment).
%most importantly it allows us to
%Here our use of   for demonstrating canonical typicality by introducing randomness on the choice of the global state and looking at the properties of the local state.
% as a purely kinematic and local 
%property, i.e. a property due to the geometry of the space of states of embedded quantum systems.
%as a purely \textit{kinematic} property: almost all pure states of large composite systems are locally canonical. 
 %Here, we present \textit{dynamical} and \textit{local} results: 
 We show that for almost all interaction Hamiltonians (in a sense to be defined hereafter) the reduced density matrix of the subsystem
has a \textit{typical} dynamics. In other words, the microscopic structure of interaction Hamiltonians does not matter 
and reduced density matrices have a self-averaging property \textit{at all times}.
These results have two important consequences: first they can explain the absence of sensitivity to microscopic details of processes like for instance thermalisation.
%when considered as a dynamical and \textit{relative} process (i.e. relevant to a subsystem coupled to quantum environment) .
Second they provide the rigorous ground for an averaging procedure over random interactions which can be used for analytical calculations performed with full generality i.e. for \textit{arbitrary} system, environment, and initial state (see
 \cite{IthierTypEqState} for an application to the equilibrium state of an embedded quantum system).
 % by averaging over some randomness introduced \textit{only} at the level of the interaction Hamiltonian.
%In a sense, our results provide a new method for handling calculations in quantum theory relying on
%the deliberate introduction of incomplete information 
%(and therefore randomness)
%about the nature of the physical interaction between a quantum system and its environment. 
More generally, this work provide a rigorous justification for a new kind of ergodicity partly envisioned in the early work of Wigner and Dyson\cite{dyson_statistical_1962}
 when modeling entire nuclear Hamiltonians using random matrices. Indeed, time averages or ensemble averages over states are not required anymore, the key concept is provided by an averaging procedure over the interaction Hamiltonian only. 
 
\emph{Model.}---
We consider a system $S$ in contact with an environment $E$  and  denote by $\mathcal{H}_s$, $\mathcal{H}_e$  their respective Hilbert spaces (see Fig.\ref{Fig2}).
The composite system $S+E$ is a closed system whose Hilbert space is the tensor product $\mathcal{H}=\mathcal{H}_s \otimes \mathcal{H}_e$ and whose total Hamiltonian $\hat{H}$ is $\hat{H}=\hat{H}_s+\hat{H}_e+\hat{W}$ where $\hat{W}$ is an interaction term.
The state of the composite system $S+E$ is described by a density matrix $\varrho(t)$ obeying the well-kown relation derived from the Schr\"odinger equation: 
%(assuming $\hat{H}_s,\hat{H}_e,\hat{W}$ do not depend on time):
$ \varrho(\tau) =\hat{U}_\tau  \varrho(0) \hat{U}_\tau^\dagger$ with $ \hat{U}_\tau=e^{- i \hat{H} \tau}$ the evolution operator and $\tau=t/\hbar$. The initial state of $S+E$: $\rho(0)$, can be chosen arbitrarily. In particular it should be noted that the environment state can be pure, there is no requirement of thermal equilibrium in sharp contrast  to most models of thermalisation and decoherence\cite{Weiss}.
As $S$ is not a closed system, its state is described by the reduced density matrix~\cite{Landau1927}:
%,NielsenChuang}: 
$\varrho_s(\tau ) = \Tr_e  \varrho(\tau),$ where $\Tr_e$ denotes the partial trace with respect to the environment.
The aim of this Letter is to characterize, in the limit of a large environment (i.e. $\dim \mathcal{H}_e \to \infty$),  the behavior of $\varrho_s(\tau)$
considered as a function $f$ of the interaction $W$:
$$ f(\hat{W})= \Tr_e \left( e^{-i \tau (H_0+W)}  \varrho(0) e^{i \tau (H_0+W)} \right), $$
 all other relevant parameters being fixed: $\tau$, initial state $\varrho(0)$, system Hamiltonian $\hat{H}_s$, environment density of states $\rho_e(\epsilon)$.
 For this purpose, we
%make the following hypothesis on the interaction Hamiltonian:
%between $S$ and the environment: 
%At the stage we consider 
 introduce \textit{deliberately} some randomness in $\hat{W}$
by assuming a lack of knowledge on the microscopic details of $\hat{W}$ (i.e. the matrix elements of $\hat{W}$ themselves) compatible with some
 symmetry property (e.g. real or hermitian symmetry) and a macroscopic constraint (here the strength $\sigma_w^2 = \Tr(\hat{W}.\hat{W}^\dagger)/ \dim \mathcal{H}$ is fixed).
 The main result of this paper shows that, surprisingly, the uncertainty on the $W_{n,m}$ does not preclude extracting useful information on the behavior of $f(\hat{W})=\varrho_s(\tau)$
  for several classes of random interactions.
 Here, we will consider $W$ to be either a Wigner Band Random Matrix (WBRM i.e. of the type $W_{i,j}= a((i-j)/b) Y_{i,j}$ where $Y$ is a Wigner Random Matrix and $a(x)$ is a deterministic band profile, ``b'' being the bandwidth) or a Randomly Rotated Matrix (RRM i.e. of the type
 $ U.D.U^\dagger $ with $D$ real diagonal fixed and $U$ unitary or orthogonal Haar distributed).
 In both cases, $W$ should be centered (i.e. $\Tr(W)=0$) and with fixed spectrum
  variance (i.e. $\Tr(\hat{W}. \hat{W}^\dagger)/\dim \mathcal{H}=\sigma_w^2$ fixed).
Such a choice for $\hat{W}$ is justified by the fact that the WBRM ensembles are attractive for modeling interactions in tight binding systems or generic conservative systems with complex behavior like heavy atoms and nuclei (see e.g. \cite{FlaumbaumStructureCompoundState1994,fyodorov_wigner_1996}, references therein and the reviews in \cite{weidenmuller_random_2009,mitchell_random_2010,Borgonovi20161}). Their band structure emerges from the finite energy range of interaction which can be seen as a consequence of some selection rules. 
On the other hand, matrices from the RRM ensembles are \textit{dense}, which is a priori not compatible with a few body nature
 of interaction. Despite this, they can be useful for modelling \textit{local} statistical properties of more physical Hamiltonians (see\cite{PorterBook,brody_random-matrix_1981,Mehta} and the discussion in\cite{Borgonovi20161}).
 At this stage, we provide analytical results on dynamical typicality for both
 these dense and sparse interaction matrix ensembles but it should be noted that our method could be used to 
 study other ensembles.

Then we consider $f(W)=\varrho_s(t)$ and notice that such a function is defined on a high dimensional input space: the set of random interactions (dimension
$ \dim^2 \mathcal{H} $)
% the set $H_{\rho_W}$, 
and takes values in a much lower dimension output space (since $ \dim \mathcal{H}_e \gg 1 $). In addition, as we will see in the following, % we show in the Supp. Info. that 
the partial trace is balancing the dependence of 
$f(\hat{W})$ on all the $W_{n,m}$ (i.e. there is no outlier $\hat{W}_{n,m}$ on which $f$ depends mostly).  As a consequence, 
%we show in Methods Sec.\ref{method:conc_of_measure} that
 the reduced density matrix will exhibit, for all random matrix ensembles considered 
before, a phenomenon known as the ``concentration of measure"\cite{MR1387624}  that we shall describe now and quantify rigorously.

\emph{Concentration of measure.}---
Central Limit Theorems (CLTs) provide the simplest illustration of this phenomenon.
 Considering for instance an experiment whose measurement output $X$ is the subject of a random error, then
  averaging the outcomes of $N$ independent measurements $\{X_1,...,X_N \}$ taken in stationary conditions will increase the signal 
 over noise ratio typically by a factor $\sqrt{N}$ (i.e. $g(X_1,...,X_N) = 1/N \sum_k X_k$ has a standard deviation $\sigma_g=\sigma_X/\sqrt{N}$). 
 Surprisingly, the decrease of the relative fluctuations is not limited to the empirical average considered above but appears under very
 general assumptions on a function $g$ and the probability distribution $P$ of its input:
  this is the so-called ``measure concentration''~\cite{MR1387624}, a phenomenon envisioned in the early work of 
P. L\'{e}vy, but formally  established by V. Milman and M. Gromov in the $1970$'s and the $1980$'s.
  %(see the celebrated paper by Talagrand~\cite{MR1387624}). 
  It can be described informally as follows: a numerical function that depends in a regular  way on many
   random independent variables, but not too much on any of them, is \emph{essentially constant} and equal to its mean value almost
    everywhere.   
The simplest analytical tool to quantify this phenomenon (and the one we use here)
 is the Poincar\'{e} inequality\cite{agz}:
 a Riemannian manifold $\Omega$ equipped with a probability measure $P$ 
 is said to verify a Poincar\'e inequality if there 
exists $C>0$  such that for all  functions $g: \Omega \rightarrow \mathbb{R}$ continuously differentiable,
 one has
\begin{equation}
\label{PoincareIneq} \sigma_g^2 = \E[ | g(X)  - \E[g] |^2 ] \leqslant  \frac{1}{C} \E[ || \vec{\nabla} g ||^2] 
\end{equation}
where
\begin{equation*}
  ||\vec{\nabla} g ||^2 =   \sum_k  \left( \frac{\partial g}{ \partial X_k} \right)^2  
\end{equation*}
 and where the average $\E$  is relative to $P$.
In other words, the variance of $g$ is controlled by the typical gradient strength over a constant $C$ (that we call the ``Poincar\'{e} constant'' in the following) usually related to the variance of the input.
 A first interesting case is when $C$ does not depend on the dimension and
  $|| \nabla g ||^2 \approx 1/N$, e.g. with the empirical average case considered above  where the central limit theorem provides 
  an equality case: $g(X_1,...,X_N)= \frac{1}{N} \sum_i X_i$, the CLT provides $\sigma_g^2 = \sigma_X^2/N $, since $||\nabla g||^2 = 1/N$ and
   $1/C=\sigma_X^2$. Another interesting case is when  $1/C \approx 1/N$ and $||\nabla g||$ is upper bounded by a 
   constant independent of the dimension, e.g. when $\Omega$ is the hypersphere $\mathbb{S}^{N}$ of radius $1$ of
  $\mathbb{R}^N$ (i.e. $\{ X \in \mathbb{R}^N / ||X||=1  \}$) equipped with the Haar measure (i.e. the isotropic probability measure).
  In both cases, the relative  fluctuations of $g$ around its mean value are  "squeezed" like $1/\sqrt{N}$ 
  %which can be very small when considering input spaces for $f$ made of Hamiltonians operating on large dimension Hilbert
  % spaces, like the typical ones of environments. 
  and the function is said to be \textit{concentrated} around its mean, which is thus a very good estimate of $g$ at any point of the space of high dimensionality.
%which is a sort of generalization of the law of large numbers.
Here, we consider the Poincar\'e inequality to be sufficient for our purpose, however it is possible to 
 go beyond and characterize the statistics of $g$ in more detail (see e.g.~\cite{MR1387624,MR856576}).

\emph{Main result.}---
The set of Hermitian matrices we consider for $W$, when endowed with the probability measures we consider here  (WBRM and RRM, see Supp. Mat. Sec. 2 for details) 
%random matrix ensembles we consider for $\hat{W}$ 
%with $\Tr(W.W^\dagger)/\text{dim} \mathcal{H}=\sigma_w^2$ fixed) 
verify Poincar\'e inequalities  with constants 
%calculated
%$U.D.U^\dagger$ with D diagonal real fixed and $U$ unitary or orthogonal Haar distributed), 
%in the Supp. Mat. Sec. 2 % \ref{PoincareConstants} and
 both lower bounded in the following way:
%the following lower bound on the Poincar\'e constant: 
$C \geq \frac{\dim \mathcal{H}_s \dim \mathcal{H}_e}{ 2 \sigma_\omega^2 } $
  where $\sigma_\omega^2= \Tr(\hat{W}.\hat{W}^\dagger)/\dim \mathcal{H}$.
%they verify a Poincar\'{e} inequality with a constant $C$ scaling typically like the dimension of the total Hilbert space
In addition, we provide in Supp. Mat. Sec.1  %\ref{SupInfoConcentration} of 
 the following upper bound on the gradient of $f(W)= \varrho_s(t)$:
%Sec.\ref{SupInfoConcentration} of the
 %the gradient of $f$ is upper bounded uniformely with the dimensin of the environment Hilbert space:
\begin{equation}
\label{UpperBoundGradient}
|| \nabla_W \varrho_s ||^2 \leqslant 2 \frac{t^2}{\hbar^2} \dim \mathcal{H}_s.
\end{equation}
Using the Poincar\'e inequality in Eq.\eqref{PoincareIneq}, we get the upper bound on the variance of 
the fluctuations of $\varrho_s$ away from its mean behavior :
\begin{equation}
\label{ConcentrationRhoMat}
\sigma_\varrho^2  = \E[|| \varrho_s -\E[\varrho_s]  ||^2] \leqslant  \frac{ 4   \sigma_\omega^2 t^2 }{\hbar^2}
\frac{1}{\dim \mathcal{H}_e} 
\end{equation}
 with $ ||A||^2= \Tr(A.A^\dagger)$ and where $\E$ is the average performed over the random matrix ensemble considered. 
 %It is important to note that there is no averaging procedure over states: the initial state is fixed once and for all.
 One should note that there is a priori no genuine physical effect responsible for the dependence of the upper bound in Eq.\eqref{ConcentrationRhoMat} with time. This upper bound is not optimal but is sufficient to demonstrate typicality:
%where the norm $||.||$ refers to the trace norm: $||A||^2= \Tr(A.A^\dagger)$.
for fixed values of 
%$\dim \mathcal{H}_s,
$ \sigma_w$ and $t$, one has $\sigma_{\varrho} \rightarrow 0$ when
 $\dim \mathcal{H}_e \to \infty$. 
The mesoscopic fluctuations of this function are decreasing down to zero as the dimension of the environment Hilbert space goes to infinity.
%A numerical example of the use of L\'evy's lemma is given in the Methods section to illustrate how
We argue that this phenomenon is at the core of irreversible processes such as thermalisation: it is responsible for a self-averaging property of the reduced density matrix of $S$, considered as a function of the 
interaction Hamiltonian. This means that $\varrho_s(t)$ has a ``typical'' behavior for almost all
interaction Hamiltonians within the classes considered. We have performed numerical simulations for a two level system coupled to a quantum environment (see Fig.\ref{Fig2}).
For small $\dim \mathcal{H}_e$, we observe a random pattern (analogous to a ``speckle'' pattern in optics) for the probabilities of occupation of the states of $S$. This pattern can be seen as a
 signature of the microscopic structure of the interaction Hamiltonian. As $\dim \mathcal{H}_e$ is increased, the amplitude of the fluctuations of this ``speckle'' pattern decreases and
  a typical behavior (i.e. independent of the details of $\hat{W}$) emerges.
%as the dimension of the environment Hilbert space increases.
To get an intuitive understanding of the concentration of measure phenomenon,
% which can be seen as a generalized Central Limit Theorem for ``reasonable'' functions, 
it is worth comparing it to the Monte Carlo method for numerically estimating the integral of a function over some subspace.
A good estimate of this integral is provided by a discrete average of the function sampled randomly over the subspace.
Measure concentration provides a path along the opposite direction: one is interested in the value of a function at 
a \textit{single} point of the subspace and in \textit{most cases}, a very good estimate of this value is the average of the function 
over the subspace. 
As a consequence, it should be noted that this phenomenon provides an approximate way of calculation of $\varrho_s(t)$ simply by averaging:
$\varrho_s(t)= \Tr_e( \varrho(t) ) \approx \mathbb{E} [\Tr_e (\varrho(t))] = \Tr_e \left( \E[\varrho(t)]\right)$, where
% the notation 
$\mathbb{E}$ is the average over the set of interaction Hamiltonians considered. This property will be used in\cite{IthierTypEqState} in order to calculate analytically the equilibrium state of an embedded quantum system when thermalisation takes place.
% and the out of equilibrium dynamics 
%of an embedded quantum system.
%\cite{OutEqDyn}.
Finally, it is important to stress that the upper bound in Eq.\eqref{UpperBoundGradient} does \textit{not} depend on the statistics of $\hat{W}$, meaning that our framework
can be adapted straight away to other classes of random matrices as soon as a lower bound on $C$ is available
for these classes. For instance, it would be very interesting to investigate embedded random matrix ensembles\cite{Flambaum1996,Flambaum2000,Kota,papenbrock_colloquium_2007} which are relevant 
when enforcing a two body nature of the interaction.

%For the first time, our result provide a rigorous justification and an explanation for such an averaging procedure.
%  this is the first time such an averaging procedure over disorder is justified.
%$H_{\rho_W}$
 %(see Sec.\ref{method:conc_of_measure} in Methods and Sec.~\ref{SupInfoConcentration} in Supp. Info. where we provide 
 %a $O(1/\dim \mathcal{H}_e) $ upper bound on the variance of the  reduced density matrix away from its typical behavior).
 % for a full justification of this averaging procedure).
 %calculating the average of the matrix elements of $ \hat{U}_t    |\phi_m \rangle  \langle \phi_p |  \hat{U}_t^\dagger $ is sufficient for
 % our purpose.
% where $\mathbb{E} $ means average over the ensembles of Hamiltonians $H_{\rho_E,d_E}$
% for each energy involved.
\begin{comment}
We are led to consider the mean of Eq.~\eqref{FourierOverlaps} which involves the fourth order moments of the overlaps:
$\E[ \lan \phi_n  | \psi_i \ran \lan \psi_i | \phi_m\ran 
 \lan \phi_p | \psi_i \ran \lan \psi_i  |\phi_q\ran]$ and the two point probability density of the spectrum of the dressed Hamiltonian $p(\lambda,\lambda')$.
 % (where $\E$ refers to the average respective to the random matrix ensemble considered).
% which is  coefficients.
% between the eigenvectors of $\hat{H}_0$ and 
 %eigenvectors of  $\hat{H}_0+\hat{W}$.
  Before we investigate these quantities, we shall justify more generally the use of random interactions for understanding quantum dynamics in general and thermalisation in particular.
\end{comment}

% Possible Exp Test
\emph{Possible experimental test.}---
In order to test experimentally this dynamical typicality property, ultra cold atoms in optical lattices seem the
most promising\cite{bloch_quantum_2012,schreiber_observation_2015}, since these systems  provide now the best level of isolation from uncontrolled degrees of freedom, and as such, enforces the required global unitary evolution over a sufficient duration.
 Such systems allow an accurate and independent control of the 
 relevant parameters of the Hamiltonian: one site interaction and inter-site hoping can be tuned conveniently by the lasers creating the lattice potentials and randomness can be inserted using optical speckle patterns. 
 %Note that such an in-situ engineering of the Hamiltonian allows to probe the thermalisation crossover (experimentally  
 %investigated for a particular Hamiltonian: Aubry-Andr\'{e} model) associated to the Many Body Localization phase transition.
In addition, these systems provide local observable measurement, like for instance local atomic density measurement,
  % using direct imaging, 
 %site imbalance and the presence of a charge density wave). 
 %Such local probes 
and  should allow to monitor the emergence of the predicted typical behavior as system size increases (i.e. the squeezing of the fluctuating pattern that we observe numerically on Fig.(\ref{Fig2})). 
% In addition non trivial density of states (e.g. with bands and gaps) can be engineer in these systems, and should lead to non thermal equilibrium states
% (see Eq.\eqref{ProbaS2}). Finally, the crossover to thermalisation, as coupling strength increases, was investigated  specifically in the context of the Aubry-Andr\'{e}
% model in~\cite{schreiber_observation_2015} and could be also tested with a random coupling  in a similar experimental setup.

\emph{Discussion and Summary.}---
 Before summarize our results, it is interesting to put them into perspective by 
 %explain in depth why we use random matrices for the interaction Hamiltonian by
  considering the motivations of the first users of random matrices in a physical context,
 %The theory of random matrices was initiated in the 1930's by the statistician Wishart
 %and further developed  by several physicists and mathematicians, including 
 Wigner and Dyson,
 %and  Mehta
 who were studying the high energy neutron scattering spectra of medium and large weight nuclei
  (see for instance the reviews in~\cite{brody_random-matrix_1981,Guhr1998,weidenmuller_random_2009,mitchell_random_2010} in references therein). 
% Let us consider the motivations of the first users of 
%random matrices in a physical context: Wigner and Dyson, who were in the 1950's and 1960's  studying (among other subjects) the
% high energy neutron scattering cross-sections of medium and large weight nuclei (see the review in~\cite{Guhr1998189}). 
%Such spectra could be measured at the time with sufficient accuracy and over a sufficient range to allow the study of its statistical  properties.
Facing the complexity of such spectra, Wigner renounces inferring the Hamiltonian of this complex $N$-body system from
 the experimental data and operates a drastic change of point of view. He assumes some \textit{statistical} hypothesis on 
 the entries of the Hamiltonian considered as a random matrix, 
which are compatible with the  general symmetry properties associated with the integrals of motion.
%However, since the cross sections could be measured with sufficient energy accuracy
%and over a sufficient wide energy range, some statistical study was possible. This led him to make the following drastic change of
% point of view
More precisely, the candidate nuclear Hamiltonian is written as a block-diagonal matrix where each block corresponds to given values
 of the ``good" quantum numbers (i.e. the conserved quantities). The entries of each block are then assumed to be independent identically 
 distributed random variables with variance and mean depending on the conserved quantities associated with the block considered.
Then averaging over this ensemble of Hamiltonians, one gets typical properties and in particular the average behavior of energy levels which is of prime importance for nuclear reactions and can be compared to experimental data.
%The idea is to make statistical hypothesis which are compatible with the symmetries of the problem.
%It is interesting to compare Wigner's method with the assumptions made in statistical physics: 
%in this framework, the observer renounces the full knowledge of the microscopic state, assuming  the only knowledge that a global 
%constraint is realized (e.g. the energy, particle number, etc... are set to some value), and then, taking a bayesian point of view, 
% assumes that all micro-states compatible with the constraint are equiprobable.
%Some thermodynamical quantities can then be calculated and the properwties of the canonical state can be established, by
% considering an ensemble of identically prepared systems all governed by the same Hamiltonian but differing in initial conditions
% (Gibbs ensemble) and averaging over the ensemble. Then by assuming an ergodic hypothesis, these theoretical predictions can
% be compared with experimental values which are time averaged quantities of a \textit{single} system.
%In standard statistical physics, the observer takes a bayesian point of view and assumes that all microscoopic states compatible 
%with a macroscopic constraint (the accessible states) are equiprobable. This allows to compute 
As noticed by Metha~\cite{Mehta} and Dyson~\cite{dyson_statistical_1962}, the approach of Wigner is much more radical
than the standard statistical physics approach: there is a subjective lack of knowledge not on the state of the system, but on the \textit{nature} of the system itself. 
%Wigner considers ensembles
%of systems driven by different microscopic (at the level of nucleon-nucleon interaction) Hamiltonians which have the same macroscopic (at the level of the whole nucleus) properties.
%He thus focuses on typical properties coming from the underlying symmetries associated with the
%conserved quantities of the evolution. 
Dyson justifies this point of view as follows~\cite{dyson_statistical_1962}: ``We picture a complex nucleus as a black box in which a 
large number of particles are interacting according to unknown laws. As in orthodox statistical mechanics we shall consider an 
ensemble of Hamiltonians, each of which could describe a different nucleus. There is a \textit{strong logical expectation}, though no
 rigorous proof, that an ensemble average will correctly describe the behaviour of one particular system which is under observation". 
The approach of Wigner and Dyson has proved to be extremely fruitful not only for explaning 
nuclear spectral fluctuations (e.g. the nuclear level spacing properties) and nuclear reactions, but also in various other fields
(e.g. mesoscopic physics, Quantum Chromo Dynamics, complex atoms and molecules, two dimensional gravity, conformal field theory, see \cite{Guhr1998} for a review).
This strongly suggests that randomness \textit{constrained by symmetries} provide a general principle for modeling not only disorded systems but also complex N body systems.
% In addition, these principles do not rely on dynamical laws.
%Aside from nuclear systems, RMT has been applied successfully to numerous physical systems with large number of degrees
% of freedom: complex atoms and molecules,  (QCD, 
%disordered quantum systems, localization in mesoscopic systems see \cite{Guhr1998189} for a review)
%  the chiral transition in QCD
The point of view we adopt in this paper is analogous to Wigner and Dyson's, however more with broader applicability
since system and environment Hamiltonians are \textit{arbitrary} and randomness is introduced
only at the level of the interaction Hamiltonian.
We considered two random matrix ensembles for the interaction %with a high level of symmetry, 
however our framework is general and can be used to study other systems with different classes of interaction Hamiltonians (e.g. conserving some set of observables
 or enforcing  the two body nature of the interaction).
But most importantly, we can justify rigorously the ``ensemble averaging" over this randomness by the phenomenon of measure concentration, and as a consequence the ``strong logical expectation'' mentioned by Dyson in the previous quotation.
Measure concentration provides a rigorous explanation for the absence of sensitivity to microscopic details of processes like for instance thermalisation:
reduced density matrices of embedded systems have the property of being self-averaging.
In the some sense, it sets the ground for what should be considered as a new kind of ergodicity:
time averages or ensemble averages 
over microscopic states are not required and can be replaced by ensemble averages over interaction Hamiltonians in order to
 obtain the typical dynamics.
 
 \emph{Acknowledgements}---
 We are indebted to D. Esteve and H. Grabert for their critical reading of the manuscript, their support and the numerous discussions and to J.M. Luck for the discussions.

\bibliographystyle{apsrev4-1}


%\bibliography{Thermalisation}     %  359 words

\begin{thebibliography}{37}
\bibitem [{Not()}]{NoteMeasConc}%
  \BibitemOpen
  \href@noop {} {}\bibinfo {note} {This phenomenon should not be confused with
  ``entanglement concentration'', a different concept specific to quantum
  information theory}\BibitemShut {NoStop}%
\end{thebibliography}


%%%%%%%%%%%%%%%%
%%  Figures
%  captions: about 240 words 
%%%%%%%%%%%%%%%%

\newpage
\begin{widetext}

\clearpage
%\newpage

%\begin{comment}
\begin{figure}
\includegraphics[width=1.1\textwidth]{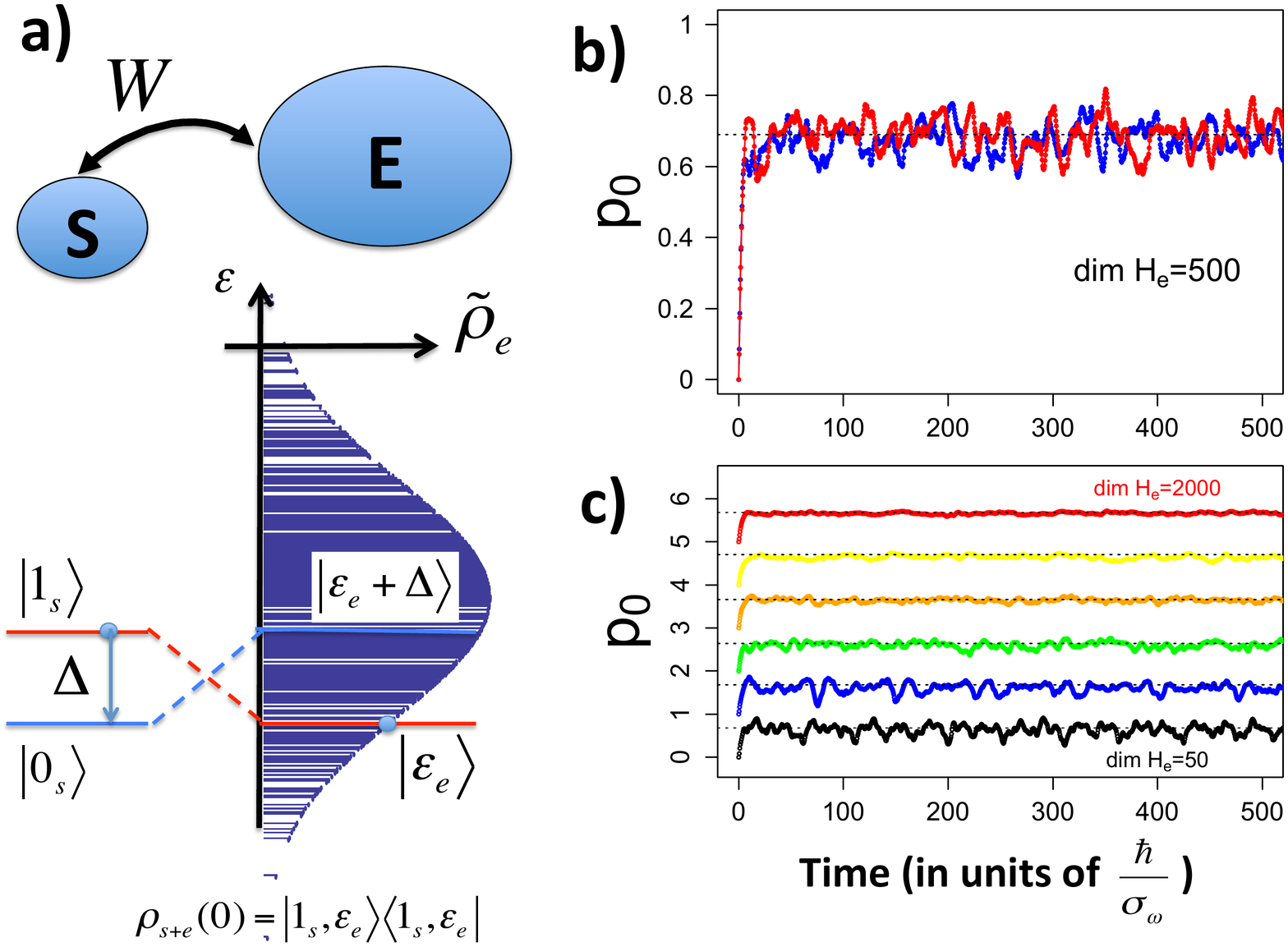}
\caption{
\textbf{Emergence of the typical dynamics.}
Numerical simulations of the time evolution of the reduced density matrix of a system $S$ coupled
 to a quantum environment.
%\textbf{Numerical simulations on a two level system.}
%the  of the time evolution of the reduced density matrix of system $S$
% a two level system $\{|0\ran,|1\ran \}$ 
%coupled to a quantum environment $E$.} 
%We consider the particular case of a two level system $|0\ran, |1\ran$ coupled to a quantum environment.
\textbf{a)} Schematic representation of a system $S$ coupled to
a large environment $E$, we consider here the particular case where $S$ is a two level system. The density of states of the environment $\tilde{\rho}_e$ is chosen to be gaussian distributed with standard deviation 
$\sigma_e=1$
and the gap of the two level system is $\Delta=1$. 
Written in the eigenbasis of $\hat{H}_s+\hat{H}_e$, the interaction $\hat{W}$ is an hermitian matrix such that 
$\{\Re(W_{i,j})\}_{i \geqslant j}$ and $\{\Im(W_{i,j}) \}_{i>j}$ are independent centered gaussians with standard deviation $\sigma_w/\sqrt{2\dim \mathcal{H}}$ (providing $\Tr(W^2)/\dim \mathcal{H} = \sigma_w^2$).
\textbf{b)} From an initial pure state of the composite system $\varrho_{s+e}(0)= | 1_s \ran \lan 1_s | \otimes | \epsilon_e \ran
\lan \epsilon_e |$ (with $\epsilon_e \approx -1.27 $ and $\dim \mathcal{H}_e=500$), 
%where $\epsilon_e$ is set to one third of the spectrum 
we numerically integrate the Schr\"odinger equation and calculate the reduced density matrix of $S$: $\varrho_s(t)=\Tr_e(\varrho_{s+e}(t))$, for two different realizations of the random coupling $W$ (red and blue, with coupling strength $\sigma_w=0.2$).
%same statistics as for Fig.~\ref{Fig2} with coupling strength $\sigma_w=0.2$).
%$|\epsilon_e \ran$ is the $200^{th}$ eigenvector by order of increasing energy.) 
 The probability for $S$ to be in its ground state $p_0= \lan 0_s | \varrho_s | 0_s \ran$ is plotted as a function of time. After a transient regime 
 %(studied in another publication), 
 a stationary regime 
takes place characterized by a ``speckle'' pattern. 
This pattern is a signature of the microscopic structure of the interaction
 Hamiltonian. The theoretical prediction (black dashed line) is provided by \cite{IthierTypEqState} where 
 an averaging procedure over $W$ (justified by measure concentration) allows to calculate analytically the typical stationary state of $S$: for the set of parameters considered here, one has $p_0 \propto \tilde{\rho}_e(\epsilon_e+\Delta)$ 
 and $p_1\propto \tilde{\rho}_e(\epsilon_e)$.
It should be noted that remarkably in this case, this prediction matches the one of the microcanonical ensemble.
 %Eq. \eqref{ProbaS2} and is a ratio of densities of states.
 \textbf{c)} The dimension of the Hilbert space of the environment is varied from $50$ (black) to $2000$ (red) to show
 the concentration of measure phenomenon (predicted by Eq.\eqref{ConcentrationRhoMat}): the amplitude of the speckle pattern decreases and a \textit{typical} asymptotic state emerges (curves are shifted vertically for clarity).}
 %in the ``mesoscopic'' fluctuations
%  and decreases in amplitude when the dimension of the environment
\label{Fig2}
\end{figure}
%\end{comment}

% 250 words in caption

%\setcounter{section}{0}
%\renewcommand*{\thesection}{Appendix \Alph{section}}

\clearpage
%\newpage

\section{Supplementary Material}

\subsection{ Concentration of the reduced density matrix of the subsystem $S$}
\label{Concentration}

To use the Poincar\'e inequality in Eq.\eqref{PoincareIneq} and show that the reduced density matrix is concentrated, we calculate here:
\begin{itemize}
\item an adequate upper bound on the variance of the gradient of $\varrho_s(t)$ considered as a function of $W$,
\item the Poincar\'e constants for the probability measures on spaces of interaction Hamiltonian we consider in this paper (WBRM and RRM).
 \end{itemize}
 As we will see, the upper bound on the gradient of $\varrho_s$ \textit{does not} depend on the statistics chosen for $W$, meaning that our framework can be 
 adapted to other classes of interaction Hamiltonians as soon as the Poincar\'e constant (or a lower bound) can be calculated for these classes.
\subsubsection{1. Upper bound on the gradient square of $\varrho_s(t)$ considered as a function of the interaction.}
\label{SupInfoConcentration}
In this section, our aim is to provide an upper bound on the norm of the gradient of $\rho_s$ considered as a function of $W$, which is uniform in the dimension.
% This upper bound is used in the Methods section to show that the reduced density matrix of $S$ is concentrated around a typical behavior.
%show that, in the large dimensionality limit, the reduced density matrix of $S$: $\Tr_e |\psi \ran \lan \psi |$ is concentrated around a typical value (the average with respect to the interaction $W$). To proceed we will first
% then we will consider the Poincar\'{e} constants of various probability measures on ensembles of interaction Hamiltonians.
%\subsubsection{Upper bound on the gradient of $\rho_s$ as a function of the interaction Hamiltonian}
Recalling the definitions
$$  | \psi \ran \lan \psi | = U_\tau \rho(0) U_\tau^\dagger
\qquad \text{and} \qquad U_\tau = e^{-i\tau (\hat{H}_0+\hat{W})},$$
with $\tau=t/\hbar$, we use the well known formula for the differential of the exponential map, in order to get the differential of $|\psi \ran \lan \psi |$ with respect to the interaction $\hat{W}$:
$$
\textbf{d} |\psi \ran \lan \psi | (\delta W) =  \int_0^1  h \circ g_\alpha (\delta W) d\alpha   
$$
with 
 $$ h(A)= -i \tau  [A , |\psi\ran \lan \psi | ]  \qquad \text{and} \qquad   g_\alpha(B)= U_{\alpha \tau} \; B \; U_{\alpha \tau }^\dagger \label{Comm} 
$$
where $[,]$ denotes the commutator and $\circ$ the composition of functions. We start by the upper bound
$$ \small 
|| \nabla_W \rho_s ||^2  =\left\Vert  \int_0^1  \Tr_e h \circ g_\alpha(\cdot) d\alpha \right\Vert^2  \leqslant  \int_0^1 || \Tr_e h \circ g_\alpha(\cdot)||^2 d\alpha $$
where the notation $||\cdot ||$ is for the norm defined on the ensemble of linear applications from the space of interaction Hamiltonians to the space of density matrices on $\mathcal{H}_s$:
$$||f||^2= \sum_{i,j} ||f(M_{i,j})||_F^2 \quad \text{where} \quad ||A||_F^2=\Tr(A.A^\dagger),$$
and
 $$  f :  H_{\dim \mathcal{H} \times \dim \mathcal{H}}(\mathbb{C}) \rightarrow  H_{\dim \mathcal{H}_s \times \dim \mathcal{H}_s}(\mathbb{C}),$$
$H_{n,n}(\mathbb{C})$ is the ensemble of hermitian matrices of size $n \times n$
and  $M_{i,j}$ is the matrix with zero everywhere except at the intersection of the $i^{th}$ line and $j^{th}$ column where it has a one.
Then we have the equality:
$$  \int_0^1 || \Tr_e h \circ g_\alpha(\cdot)||^2 d\alpha =  ||\Tr_e h(\cdot)||^2$$
since 
$$ \sum_{i,j} ||f(M_{i,j})||_F^2 = \sum_{i,j} ||f(U.M_{i,j}.U^\dagger)||_F^2 $$ for any unitary matrix $U$.
%$g_\alpha$ is a unitary change of orthonormal basis.
  To move on and make the partial trace easy, we write $|\psi \ran $ in the tensor basis $|\psi \ran= \sum_{s,e} \gamma_{s,e} \; |\epsilon_s \ran \otimes |\epsilon_e \ran$ where
$\{ \gamma_{s,e} \}_{s,e}$ is a $\dim \mathcal{H}_s \times \dim \mathcal{H}_e $ matrix.
Then we have for the matrix elements of $\Tr_e h(M_{a,b,c,d})$ where $M_{a,b,c,d}= |s_a \ran \lan  s_b| \otimes  | e_c \ran \lan e_d |$:
$$ \lan \epsilon_s| \Tr_e [M_{a,b,c,d}, | \psi \ran \lan \psi|] |\epsilon_{s'} \ran =  \gamma_{b,d} \; \gamma_{s',c}^* \;  \delta_{s,a}  - \gamma_{s,d} \; \gamma_{a,c}^* \;   \delta_{b,s'} $$
Taking the square modulus and summing over $a,b,c,d$:
\begin{eqnarray}
\small \nonumber
\sum_{a,b,c,d}  | \lan \epsilon_s | \Tr_e ( [ M_{a,b,c,d},|\psi \ran \lan \psi | ] ) | \epsilon_{s'} \ran |^2 =
%   \nonumber
 \sum_{b\neq s',c,d} | \gamma_{b,d}  |^2 | \gamma_{s',c}|^2 \nonumber \\
  + \sum_{a\neq s,c,d} | \gamma_{s,d}  |^2 |\gamma_{a,c}|^2  \nonumber 
 + \sum_{c,d}  | \gamma_{s',d} \gamma_{s',c} - \gamma_{s,d} \gamma_{s,c} |^2 \nonumber  \\
 = 
   \left(  \sum_{b,c,d} | \gamma_{b,d}  |^2 | \gamma_{s',c}|^2 + \sum_{a,c,d} | \gamma_{s,d}  |^2 |\gamma_{a,c}|^2 \right) \nonumber\\
  - 2  \sum_{c,d}\Re\left(  \gamma_{s',d} \; \gamma_{s',c} \; \gamma_{s,d}^* \; \gamma_{s,c}^* \right)  \nonumber
 \end{eqnarray}
then summing over $s$ and $s'$, we get the square of the norm we are looking for:
\begin{eqnarray}
\nonumber
||\Tr_e h(\cdot) ||^2 & =&  \tau^2 \sum_{s,s'} \sum_{a,b,c,d}  \left| \lan \epsilon_s | \Tr_e ( [ M_{a,b,c,d},|\psi \ran \lan \psi | ] )  | \epsilon_{s'} \ran \right|^2 
   \\ \nonumber
& = & 2 \tau^2 \dim \mathcal{H}_s -   2\tau^2 \sum_{c,d,s,s'}\Re\left(  \gamma_{s',d} \gamma_{s',c} \gamma_{s,d}^* \gamma_{s,c}^* \right)
 \end{eqnarray}
 since the normalization condition provides $ \Tr(\gamma.\gamma^\dagger)  = 1$. The second term on the right hand side is:

$$\sum_{c,d,s,s'} \left(  \gamma_{s',d} \gamma_{s',c} \gamma_{s,d}^* \gamma_{s,c}^* \right)= \Tr((\gamma.\gamma^\dagger)^2) \geqslant 0$$

We finally get that 

$$|| \nabla_W \rho_s ||^2 \leqslant || \Tr_e h (\cdot)  ||^2  \leqslant 2 \tau^2 \dim \mathcal{H}_s.$$

%Note that as the (full) trace of a commutator is always zero, it is not surprising  that the partial trace might be small provided one is tracing out over a Hilbert space
% (here $\mathcal{H}_e$) of much larger dimension than the output space (here $\mathcal{H}_s$).  

\subsubsection{2. Poincar\'e constants for various probability measures on spaces of interaction Hamiltonians}
\label{PoincareConstants}
The Poincar\'e constants are well known for the following probability measures:

\begin{itemize}

\begin{comment}
\item Ensembles of $N \times N$ Random Hermitian matrices. 
Let us assume that the laws (which do not have to be identical) of the independent entries $\{ W_{i,j} \}$ $1\leq i \leq j \leq N$ all satisfy a P.I. with constant bounded below by $Nm$ ($m$ fixed).
%Assuming that the law of each independent entry $\{ W_{i,j} \}_{1 \leq i \leq j \leq N }$ (which do not have to be identical) satisfy 
%a P.I. with constant bounded below by $N m$. 
Then the law of the matrix ensemble obtained from the tensor product of the probability measures of the entries will also verify a P.I. with the same constant $Nm$: this is due to the ``tensorizing'' property of the P.I.
In addition, considering a random variable $X$ satisfying a P.I. with constant $m$,  the variable $\alpha X$ (with $\alpha$ fixed) will satisfy 
a P.I. with constant $m/\alpha^2$.
From this, we deduce that a random hermitian matrix with a variance profile, i.e. $W_{i,j} = a_{i,j} Y_{i,j}$ with 
$Y_{i,j}  $ a Wigner matrix with Poincar\'e constant $m$ and $0\leq a_{i,j} \leq 1$ deterministic functions of $i,j$ will 
verify a P.I.  with constant bounded below by the constant of the $Y$ matrix ensemble.
This can be applied to WBRM matrix ensembles wi
\end{comment}
%TBRI ensembles: sparcity and intrinsic correlation, bandlike structur diagonal plus sparse bandlike structure (Wigner matrices with a high level of sparcity).

\item $N \times N$ Wigner Random Band Matrices.

As a start, let us consider matrices with independent real centered Gaussian distributed entries.
The Poincar\'e constant  of the Gaussian probability measure $P$ of variance $\sigma^2$ on $\mathbb{R}$ is $1/\sigma^2$. 
It has the property of tensorizing: the Poincar\'e constant for the probability measure $P^{\otimes N}$ defined on $\mathbb{R}^N$ 
is the same: $1/\sigma^2$. The complex case is similar.
As the entries of the matrices we consider have a typical variance $\sigma_w^2/N$ (to ensure $\tr(\hat{W}^2)=\sigma_w^2$), the Poincar\'{e} constant will scale like $ N/\sigma_w^2$.
In the case of more general Wigner matrices: if the entries of a random matrix are independent and all satisfy a Poincar\'{e} inequality with the same constant, then the duly renormalized (i.e. divided by $\sqrt{N}$ ) matrix also satisfies, globally, a Poincar\'{e} inequality with constant in  the scale $N$  (see Section 4.4.1 in~\cite{agz}). In addition,
 it is well known that if a random variable $X$ satisfies a P.I. with constant $m$, then the variable $\alpha X$ (with $\alpha$ fixed) will satisfy 
a P.I. with constant $m/\alpha^2$. From this property, we conclude that WBRM matrices, i.e. with entries having a variance profile of the type $W_{i,j} = a_{i,j} Y_{i,j}$ with 
 $0 \leq a_{i,j} \leq 1$ deterministic  and $Y$ a Wigner matrix (with Poincar\'e constant $ \geq m' N$, $m'$ fixed), 
 %satisfying a PI with constant scaling like $N$, 
 will also satisfy a P.I. inequality with a constant  $\geq m' N$.  % (i.e. some entries inside the band are zero).
% Note that 
 %scaling like $N$ as soon as the matrix $Y$ does satisfy a PI with constant  scaling like $N$.
 %satisfying a P.I. with 
 %constant $Nm$ ($m$ fixed)
%This can be applied to WBRM matrix ensembles.
%This can be extended to the Hermitian case.
%Considering the ensemble of hermitian matrices with independent centered Gaussian entries each of variance $\sigma^2/N$, 
%we see that the Poincar\'e constant for this probability measure is $N/\sigma^2$. 

\begin{comment}
\item more generally, Wigner matrices (see Introduction in~\cite{agz}) have the same tensorizing property: if the probability distribution of each independent entry of the matrix satisfies a Poincar\'{e} inequality with constant $C$, then the probability distribution of the whole matrix will satisfy a Poincar\'{e} inequality with the same constant $C$.
\end{comment}

\item Ensemble of $N \times N$ Randomly Rotated Matrices, i.e. of the type $\{U.D.U^\dagger \}$ where $D$ is a fixed diagonal matrix and $U$ is unitary or orthogonal Haar distributed random
matrix.

 The Poincar\'e constant $C$ is actually related to the Ricci curvature of the ensemble considered as a manifold and the variance of the spectrum of $D$ (see 
appendix F. in~\cite{agz} and the results due to Gromov):
$$C \geq \frac{N}{ 2 \sigma_D^2 } $$
where $ \sigma_D^2 =  \Tr(D^2)/N$ is the variance of the spectrum of $D$ (which is assumed to be fixed).
%Does it generalize easily for the orthogonal and quaternionic group?

\end{itemize}

To summarize: in all cases, because the variance of the spectrum $\hat{W}$ is set to a fixed value independent of the dimension $N$, the Poincar\'e constant of the probability measure of the matrix ensemble considered is lower bounded by $m N$ with $m$ fixed.

\end{widetext}

\end{document}